\documentclass[letterpaper,prl,twocolumn,showpacs]{revtex4}
\RequirePackage{graphicx}
\newcommand{\D}{\mathrm d}
\newcommand{\E}{\mathrm e}
\newcommand{\I}{\mathrm i}

\begin{document}

\title{Simultaneous numerical simulation of direct and inverse cascades in wave turbulence.}

\date{\today}

\author{A.\,O.~Korotkevich}
\email{kao@itp.ac.ru}
\affiliation{L.\,D.~Landau Institute for Theoretical Physics RAS, 2 Kosygin Str., Moscow, 119334, Russian Federation}


\begin{abstract}
Results of direct numerical simulation of isotropic turbulence of surface gravity
waves in the framework of Hamiltonian equations are presented.
For the first time simultaneous formation of both direct and inverse cascades was
observed in the framework of primordial dynamical equations. At the same time, strong long waves background was developed. It was shown, that obtained Kolmogorov spectra
are very sensitive to the presence of this condensate. Such situation has to be typical for experimental wave
tanks, flumes, and small lakes.
\end{abstract}

\maketitle
{\it --Introduction --}
In this year we have a 50th anniversary of the famous work by Phillips~\cite{Phillips1958} which was, probably, the
first attempt to give an explanation for power-like spectra of surface gravity waves observed in
numerous experiments. In recent works~\cite{Kuznetsov2004,NZ2008} the physical explanation given by Phillips was corrected.
During less than ten years after that the statistical theory of surface waves was founded: Hasselmann derived
kinetic equation for waves~\cite{Hasselmann1962}, Zakharov created theory of wave (or weak)
turbulence~\cite{ZakharovPhD,ZFL1992}, which describes solutions of this equation. Stationary Kolmogorov solutions of the kinetic equation corresponding to flux of
energy
from large to small scales (direct cascade) and flux of wave action (waves ``numbers'') from small
to large scales (inverse cascade) were found~\cite{ZF1967,ZFL1992}. This opened a way to creation of the effective
tool for
waves forecasting. The conjectures under which the theory of weak turbulence was derived includes
Gaussian statistics for waves field and resonant interactions prevalence~\cite{ZFL1992}. They are subject for
confirmation. Modern numerical methods allow to perform wave field modeling in the framework
of kinetic equation faster than real processes in nature.
At the same time, it is impossible to create waves forecasting model based on direct numerical simulation of dynamic equations. Even more, we do not need
to know velocity and elevation at every point of the surface. Statistics, especially mean wave hight and speed,
this is what really matters for estimation of operational conditions of oil platforms and cargo ships.
And this is exactly the subject of theory of weak turbulence.
In means that the problem of confirmation and correction of the waves turbulence is of great practical importance.

Experiments in the open sea and on the Great Lakes gave temporal and space spectra consistent
with the theory~\cite{Toba1973,Donelan1985,Hwang2000}.
A comprehensive review of experiments and comparison with the theory of the weak turbulence can be found in~\cite{BPRZ2005,Zakharov2005,BBRZ2007}.
Most of these experiments were performed with wind
pumping, broad in spectrum. Narrow in spectrum pumping can be realized in wave tanks or flumes. Results obtained on such state of the art devices frequently
contradict with predictions of the theory of wave turbulence. For example, in the recent
experiments~\cite{FFL2007,Lukaschuk2007} observed spectra were changing slope with variation of steepness and pumping force.

May be the most promising way to check conjectures of the waves turbulence theory is a numerical experiment.
In the case of direct numerical simulation we have the
highest possible control on the parameters of experiments and all information about the wave field.
At the same time all this data is given at the cost of the enormous computational complexity. Fast growth of
computational power and development of computational algorithms allowed direct numerical simulation of
the surface gravity waves, starting from the simulations of the swell evolution~\cite{Tanaka2001,Onorato2002,Yokoyama2004,ZKPD2005,ZKPR2007,KPRZ2008}
to the isotropic turbulence simulation~\cite{DKZ2003grav,DKZ2004,LNP2006,AS2006}. There is a hope, that this approach
together with confirmation of conjectures of the weak turbulent theory will allow us to explain phenomena observed
in experimental wave tanks.

At the same time, theory of the wave turbulence is still under development. To close the circle the recent paper by
Newell and Zakharov~\cite{NZ2008} gave second life to the Phillips spectrum, although from completely different
point of view: Phillips spectrum considered to be a solution which give a balance of transfer of energy due to
nonlinear waves interaction transfer of energy and transfer due to intermittent events, like wave breaking and
white capping.

This Letter was inspired by several recent papers. In the first one~\cite{AS2006} numerical simulation of the
isotropic turbulence with observed formation of inverse cascade was performed in the framework of Zakharov equations~\cite{ZFL1992}.
A little bit later
a group of authors~\cite{Chertkov2007} during simulation of 2D hydrodynamics observed formation of large scale
structure due to Kraichnan's inverse cascade and explored its influence on the system. Approximately at the same
time state of the art surface waves wave
flume experiment was performed~\cite{Lukaschuk2007}. Observed spectra differed from the theory wave turbulence.
Author performed a direct numerical simulation of isotropic turbulence of surface gravity
waves in the framework of Hamiltonian equations. The isotropic turbulence is a classical setup for
turbulence investigation. In nature such such wave field is usually observed in the regions with large amount of
floating broken ice. For the first time the formation of both direct and inverse cascades was
observed in the framework of primordial dynamical equations. At the same time, strong long waves background was developed. This phenomenon of ``condensation'' of
waves (following analogy with Bose-Einstein condensation in condensed matter physics) was predicted by the theory
of weak turbulence. It was shown, that obtained Kolmogorov spectra are very sensitive to the presence of the
condensate. Such situation have to be typical for experimental wave tanks, flumes and small lakes. Obtained results
can be considered as the first observation of generalized Phillips spectra, introduced in~\cite{NZ2008}
and explain some deviations from the waves turbulence theory in recent wave tank experiments.

{\it --Theoretical background --}
We consider a potential flow of ideal incompressible fluid. System is described in terms of weakly nonlinear
equations~\cite{ZFL1992,DKZ2004} for surface elevation $\eta(\vec r, t)$ and velocity potential at the surface $\psi(\vec r,t)$ ($\vec r = \overrightarrow{(x,y)}$)
\begin{eqnarray}
\dot \eta = \hat k  \psi - (\nabla (\eta \nabla \psi)) - \hat k  [\eta \hat k  \psi] +\nonumber\\
+ \hat k (\eta \hat k  [\eta \hat k  \psi]) + \frac{1}{2} \Delta [\eta^2 \hat k \psi] + 
\frac{1}{2} \hat k [\eta^2 \Delta\psi] + \widehat F^{-1} [\gamma_k \eta_k],\nonumber\\
\dot \psi = - g\eta - \frac{1}{2}\left[ (\nabla \psi)^2 - (\hat k \psi)^2 \right] - \label{eta_psi_system}\\
- [\hat k  \psi] \hat k  [\eta \hat k  \psi] - [\eta \hat k  \psi]\Delta\psi + \widehat F^{-1} [\gamma_k \psi_k] + P_{\vec r}.\nonumber
\end{eqnarray}
Here dot means time-derivative, $\Delta$ --- Laplace operator, $\hat k$ is a linear integral operator
$\left(\hat k =\sqrt{-\Delta}\right)$, $\widehat F^{-1}$ is an inverse Fourier transform, $\gamma_k$ is a
dissipation rate (according to recent work~\cite{DDZ2008} it has to be included in both equations), which
corresponds to viscosity
on small scales and, if needed, "artificial" damping on large scales.
$P_{\vec r}$ is the driving term which simulates pumping on large scales (for example, due to wind).
In the $k$-space supports of $\gamma_{k}$ and $P_{\vec k}$ are separated by the inertial
interval, where the Kolmogorov-type solution can be recognized.
These equations were derived as a results of Hamiltonian expansion in terms of $\hat k \eta$.
From physical point of view
$\hat k$-operator is close to derivative, so we expand in powers of slope of the surface. In most of
experimental observations average slope of the open sea surface $\mu$ is of the order of $0.1$, so
such expansion is very reasonable.

In the case of statistical description of the wave field, Hasselmann kinetic equation~\cite{Hasselmann1962} for
the distribution of the wave action $n(k,t)=\langle|a_{\vec{k}}(t)|^2\rangle$ is used. Here
\begin{equation}
a_{\vec k} = \sqrt \frac{\omega_k}{2k} \eta_{\vec k} + \I \sqrt \frac{k}{2\omega_k} \psi_{\vec k},
\end{equation}
are complex normal variables. For gravity waves $\omega_k = \sqrt{gk}$.

From the theory of weak turbulence~\cite{ZFL1992}, besides equipartions (Rayleigh-Jeans) spectrum,
we know two stationary solutions~\cite{ZF1967, ZakharovPhD} of the kinetic equation in the case of four-waves interaction:
\begin{equation}
\label{direct_inverse_cascade}
n_k^{(1)} = C_1 P^{1/3} k^{-\frac{2\beta}{3} - d},\;\; n_k^{(2)} = C_2 Q^{1/3} k^{-\frac{2\beta - \alpha}{3} - d}.
\end{equation}
For surface gravity waves, a coefficient of homogeneity of nonlinear interaction matrix element $\beta=3$,
the power of dispersion law $\alpha=1/2$, and the dimension of the surface $d=2$. As a result we get
\begin{equation}
\label{weak_turbulent_exponents}
n_k^{(1)} = C_1 P^{1/3} k^{-4},\;\;\; n_k^{(2)} = C_2 Q^{1/3} k^{-23/6}.
\end{equation}
The first solution $n_k^{(1)}$ describes direct cascade of energy from large pumping to small dissipative scales.
The second solution $n_k^{(2)}$ describes inverse cascade of action (or ``number'' of waves) from
small pumping to larger scales.

{\it --Numerical simulation --}
We simulated primordial dynamical equations (\ref{eta_psi_system}) in a periodic spatial domain $2\pi\times 2\pi$.
Main part of the simulations was performed on a grid consisting of $1024\times 1024$ knots. Also we performed long
time simulation on the grid $256\times 256$.
The used numerical code was verified in~\cite{DKZ2003cap, DKZ2003grav, DKZ2004, ZKPD2005, ZKPR2007, KPRZ2008}.
Gravity acceleration was $g=1$. Pseudo-viscous damping coefficient had the following form
\begin{equation}
\gamma_{k} = \cases{
0, k \le k_d,\cr
- \gamma_0 (k - k_d)^2, k > k_d,\cr}
\end{equation}
where $k_d = 256$ and $\gamma_{0,1024} = 2.7\times10^{4}$ for the grid $1024\times 1024$ and $k_d = 64$
and $\gamma_{0,256} = 2.4\times10^{2}$ for the smaller grid $256\times 256$. Pumping was an isotropic driving
force narrow in wavenumbers space with random phase:
\begin{equation}
P_{\vec k} = f_k \E^{\I R_{\vec k} (t)}, f_k = \cases{
4 F_0 \frac{(k-k_{p1})(k_{p2}-k)}{(k_{p2} - k_{p1})^2},\cr
0 - \mathrm{if}\; k < k_{p1}\;\mathrm{or}\; k > k_{p2};\cr}
\end{equation}
here $k_{p1}=28,\; k_{p2}=32$ and $F_0 = 1.5\times 10^{-5}$; $R_{\vec k}(t)$ was uniformly distributed random
number in the interval $(0,2\pi]$ for each $\vec k$ and $t$. Initial condition was low amplitude noise in all
harmonics. Time steps were $\Delta t_{1024} = 6.7\times 10^{-4}$ and $\Delta t_{256} = 5.0\times 10^{-3}$.
We used Fourier series in the following form:
\begin{eqnarray*}
\eta_{\vec k} = \widehat F[\eta_{\vec r}] = \frac{1}{(2\pi)^2}\int\limits_{0}^{2\pi}\int\limits_{0}^{2\pi}
\eta_{\vec r} \E^{\I \vec k\vec r}\D^2 r,\\
\eta_{\vec r} = \widehat F^{-1}[\eta_{\vec r}] = \sum\limits_{-N_x/2}^{N_x/2}\sum\limits_{-N_y/2}^{N_y/2}
\eta_{\vec k} \E^{-\I \vec k\vec r},
\end{eqnarray*}
here $N_x$, $N_y$ --- are number of Fourier modes in $x$ and $y$ directions.

\begin{figure}[htb]
\centering
\includegraphics[width=3.5in]{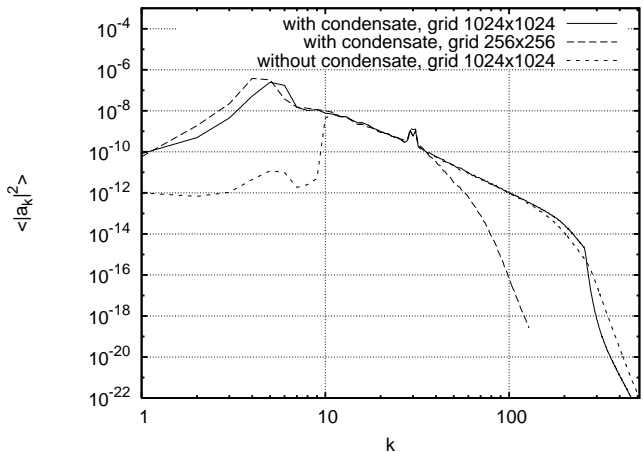}
\caption{\label{Spectra_all}Spectra $<|a_k|^2>$. With condensate on the $1024\times 1024$ grid (solid); on the $256\times 256$ grid with more developed condensate (long dashes); without condensate on the $1024\times 1024$ grid (short dashes).}
\end{figure}

\begin{figure}[htb]
\centering
\includegraphics[width=3.5in]{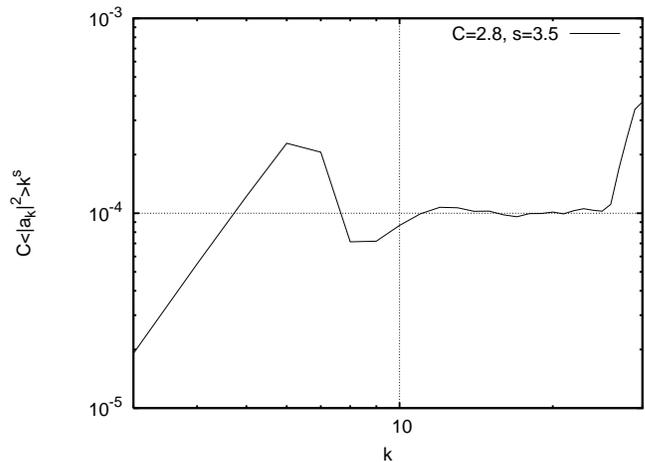}
\caption{\label{Compensated_spectra_low}Compensated inverse cascade spectra $C\langle |a_k|^2\rangle k^s$.}
\end{figure}

\begin{figure*}[htb]
\centering
\includegraphics[width=3.5in]{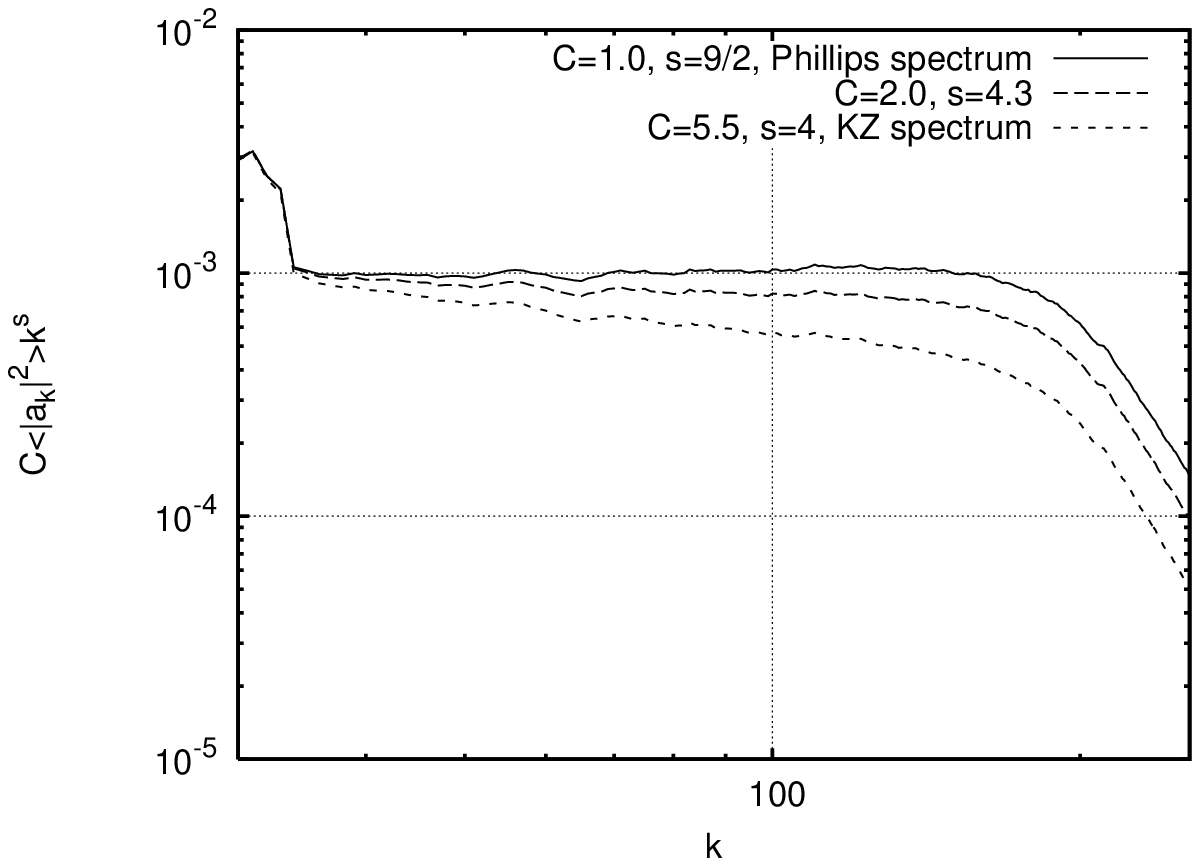}
\includegraphics[width=3.5in]{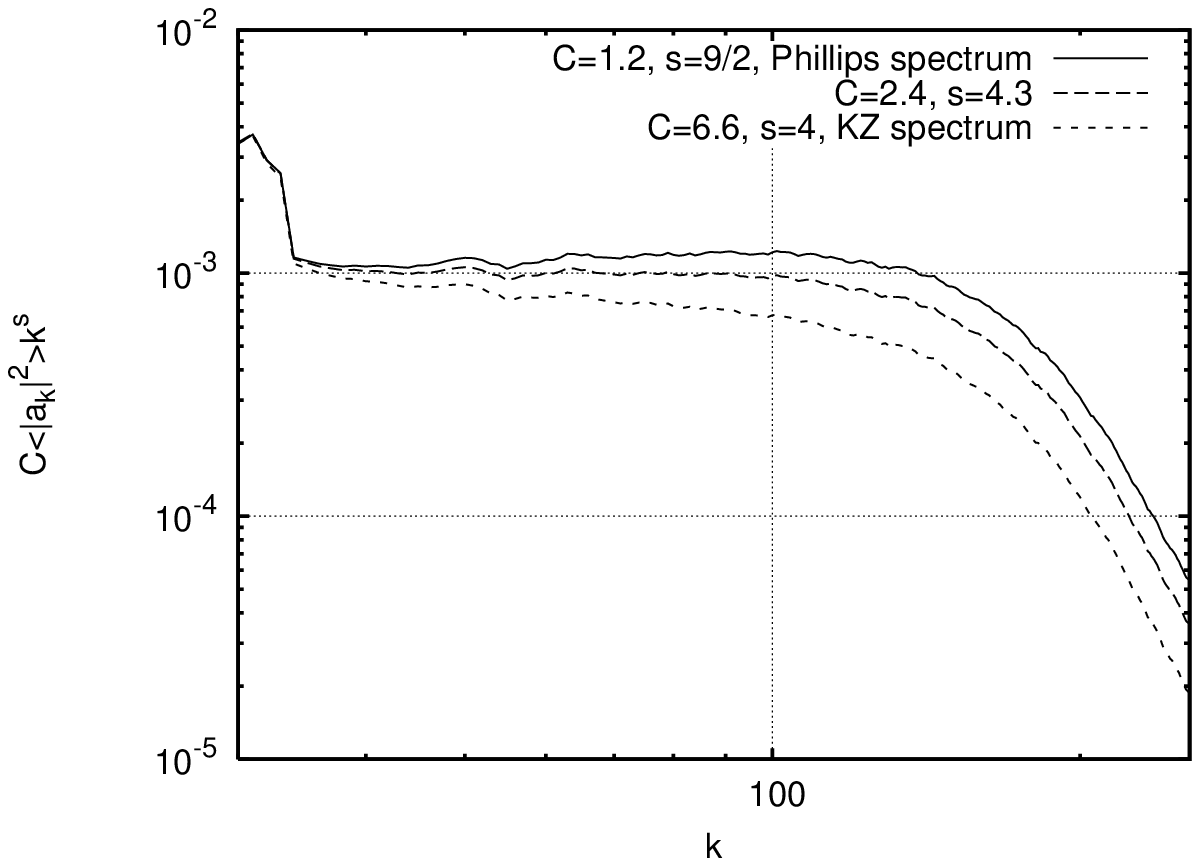}
\caption{\label{Compensated_spectra}Compensated direct cascade spectra $C\langle |a_k|^2\rangle k^s$ with (left) and without (right) condensate.}
\end{figure*}

As a results of simulation we observed formation of both direct and inverse cascades (Fig.~\ref{Spectra_all}, solid line),
although exponents of power-like spectra were different from weak turbulent solutions
(\ref{weak_turbulent_exponents}). What is important, development of inverse cascade spectrum was arrested by
discreteness of wavenumbers grid in agreement with~\cite{DKZ2003cap, ZKPD2005, Nazarenko2006}. After that large scale
condensate started to form. As one can see, value of wave action $|a_k|^2$ at the condensate region is more than
order of magnitude larger than for closest harmonic of inverse cascade. Dynamics of large scales
became extremely slow after this point. We managed to achieve downshift of condensate peak for one step of
wavenumbers grid during long time simulation on a small grid $256\times 256$
(Fig.~\ref{Spectra_all}, line with long dashes). As one can see we observed elongation of inverse cascade interval
without significant change of the slope. Unfortunately, inertial interval for inverse cascade is too short to
exclude possible influence of pumping and condensate.
We can try to estimate exponent by compensation of the
spectra in a double logarithmic scale (Fig.~\ref{Compensated_spectra_low}). The observed spectrum $\sim k^{-3.5}$
is close to weak turbulent solution (\ref{direct_inverse_cascade}). Slightly lower exponent could be explained by
weakening of resonant nonlinear interactions on the rough wavenumbers grid, which effectively decreases
the homogeneity coefficient $\beta$ in the expression (\ref{direct_inverse_cascade}).
For direct cascade spectra we also used compensation in double logarithmic scale. Results are present in
Fig.~\ref{Compensated_spectra} (left). Formally, in this case we have quite long inertial interval $32 < k < 256$,
but in reality damping has an influence on the spectrum approximately up to $k\simeq 180$. Still in this
case we have more than half of a decade. Theory of weak turbulence gives us dependence $\sim k^{-4}$
(\ref{direct_inverse_cascade}), known as Kolmogorov-Zakharov spectrum. Nevertheless, one can see that we observe
$k^{-9/2}$, known as Phillips~\cite{Phillips1958,NZ2008} spectrum.
So we need to understand, what is the reason of different spectrum slope?
What changes weak turbulent theory in this case?

To answer these questions let us compare our situation with previous works on decaying~\cite{Onorato2002, ZKPR2007,
KPRZ2008} or isotropic~\cite{DKZ2003grav, DKZ2004, LNP2006} turbulence. Immediately we have an answer: condensate
and inverse cascade spectrum! The inverse cascade's part of the spectrum is described by the theory of weak
turbulence, so let us concentrate on the strong long ($k \simeq 5$) waves' influence on much shorter waves
($32 < k < 180$), corresponding to direct cascade. We suppressed condensate by including ``artificial'' dissipation
on large scales ($k < 10$). Resulting spectrum is given in Fig.~\ref{Spectra_all} (line with short dashes).
The best wave to see the difference in characteristic waves' scale is to have a look at the surface with and without
condensate (Fig.~Surface).
\begin{figure*}[htb]
\centering
\includegraphics[width=3.5in]{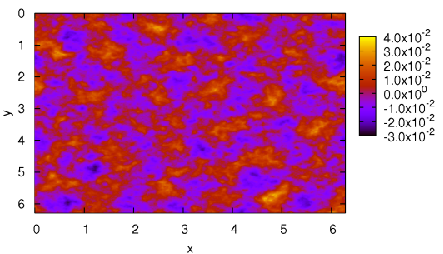}
\includegraphics[width=3.5in]{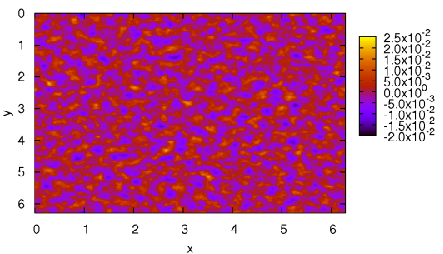}
\caption{\label{Surface}Surface of the fluid $\eta(\vec r)$ with (left) and without (right) condensate.}
\end{figure*}
Compensated spectrum for direct cascade is given in Fig.~\ref{Compensated_spectra} (right). As one can
see exponent of the spectrum changed and is now closer to the results of weak turbulent theory. The light
difference may be a result of the influence of the left edge of inverse cascade, which can play a role of
condensate for short scales corresponding to the direct cascade.

\begin{figure*}[htb]
\centering
\includegraphics[width=3.5in]{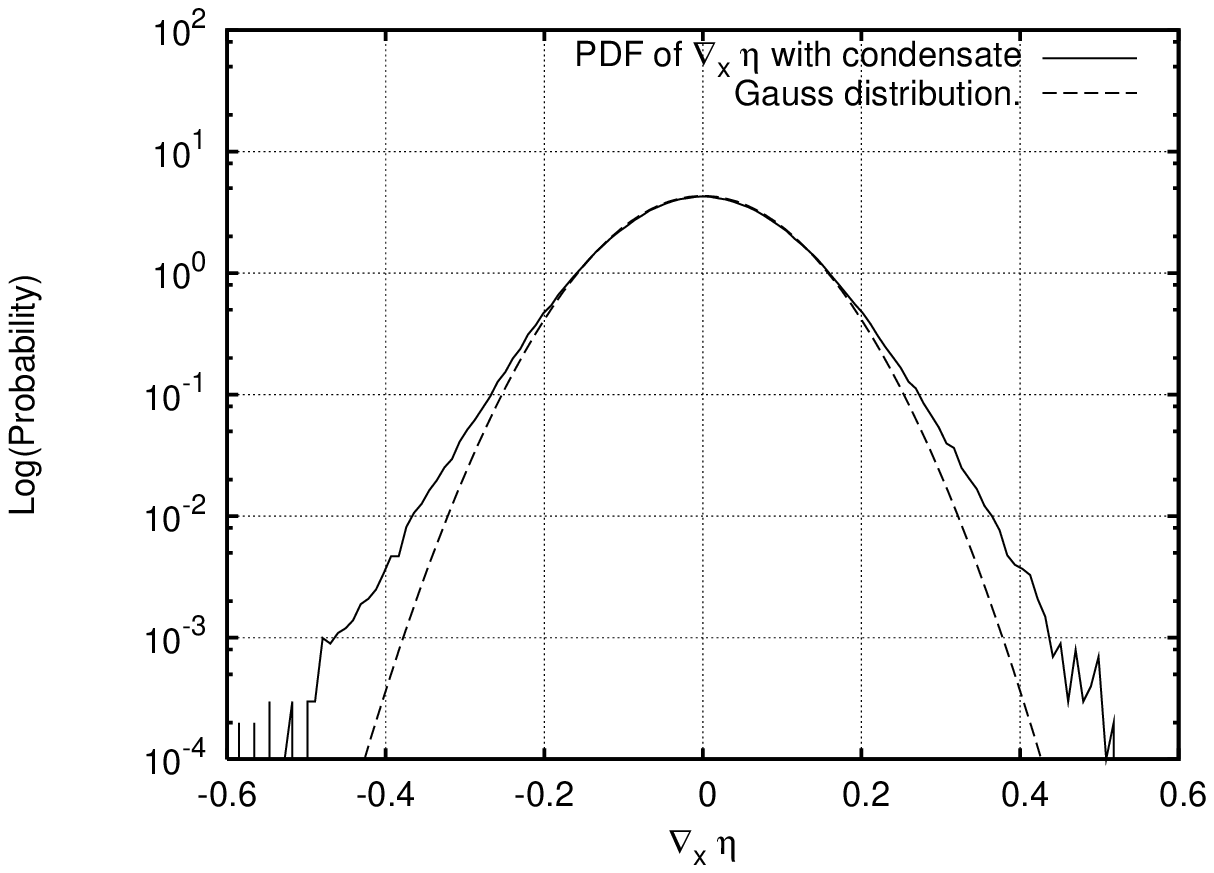}
\includegraphics[width=3.5in]{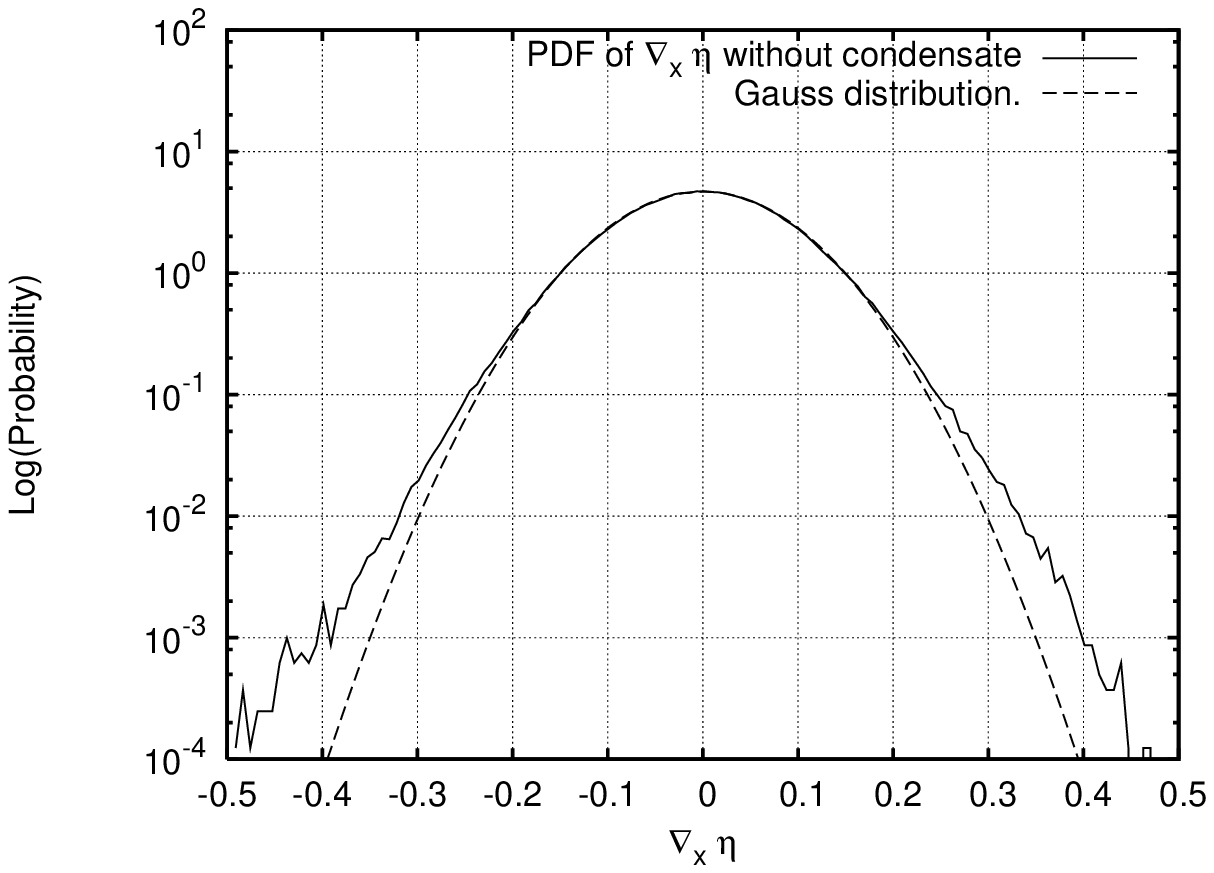}
\caption{\label{PDF_grad_eta_x}PDFs of $\vec \nabla_x \eta$ with (left) and without (right) condensate.}
\end{figure*}

\begin{figure}[tb]
\centering
\includegraphics[width=3.5in]{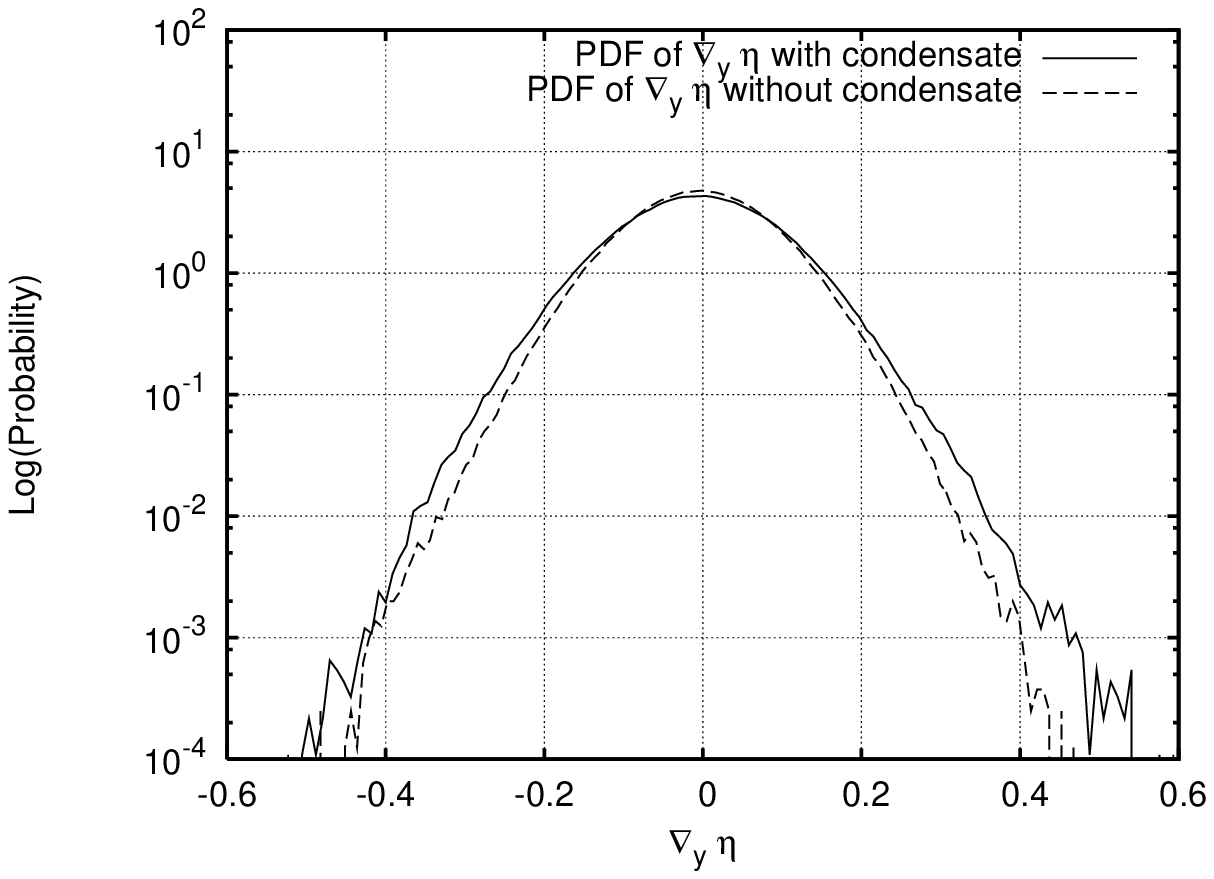}
\caption{\label{PDF_grad_eta_y_both}PDFs of $\vec \nabla_y \eta$ with (solid) and without (dashed) condensate.}
\end{figure}

\begin{figure*}[htb]
\centering
\includegraphics[width=3.5in]{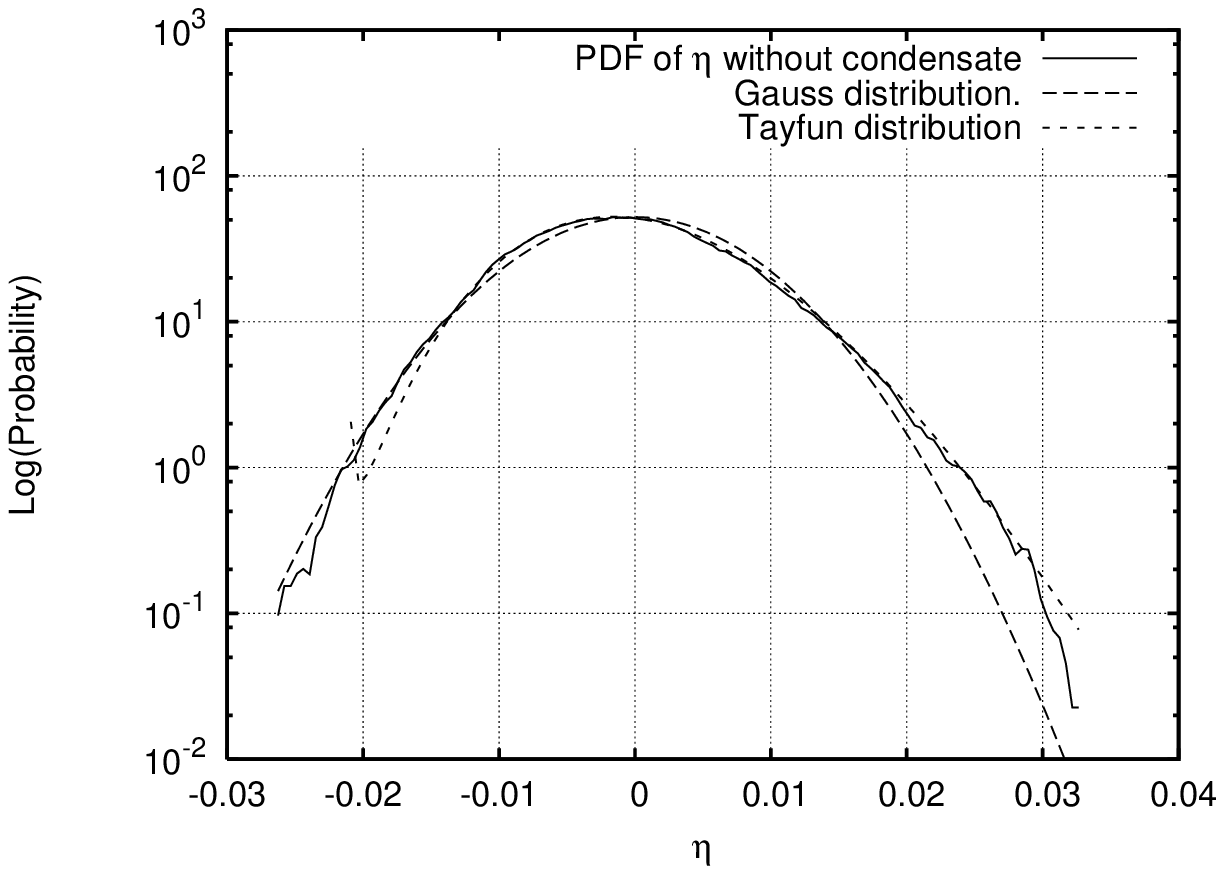}
\includegraphics[width=3.5in]{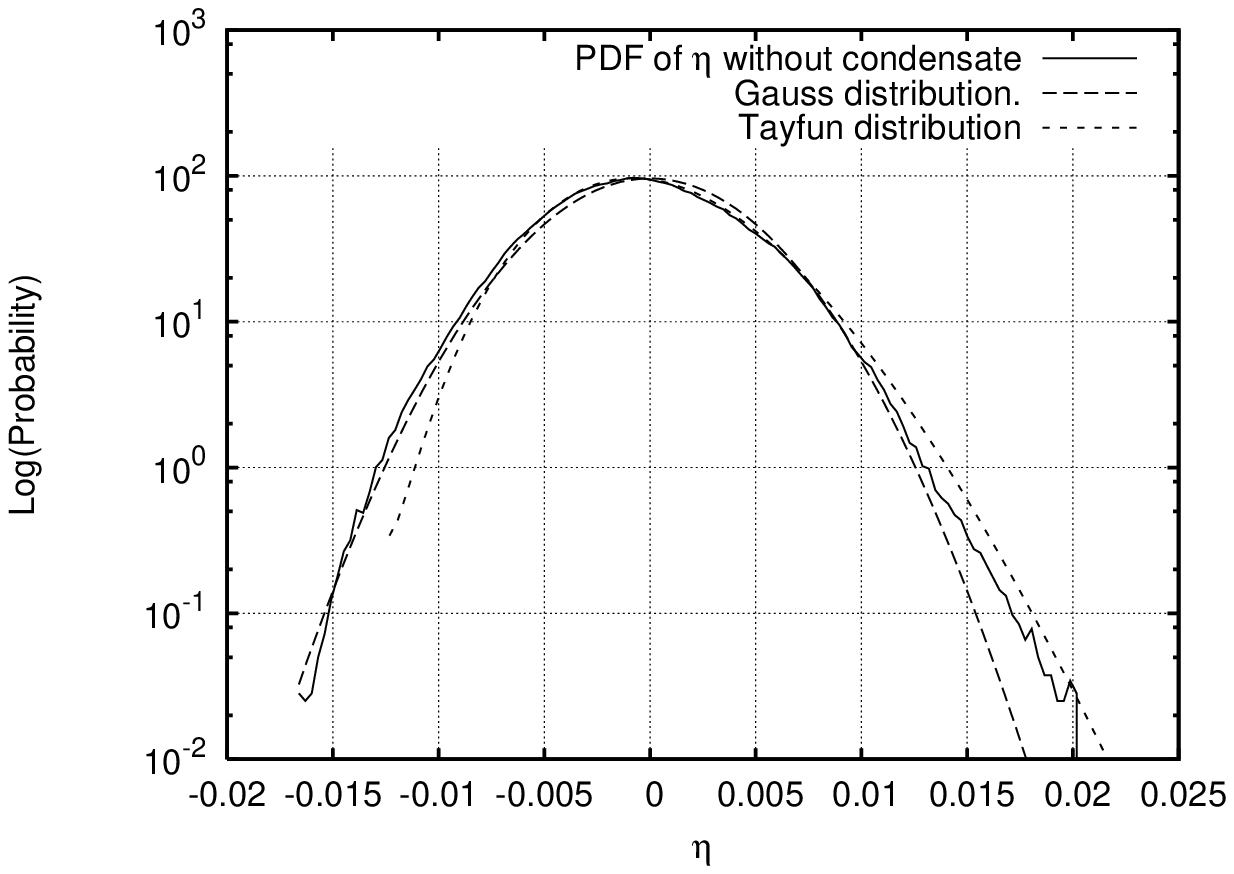}
\caption{\label{PDF_eta}PDFs of surface elevation $\eta(\vec r)$ with (left) and without (right) condensate.}
\end{figure*}

Qualitative explanation of the condensate's influence on the short waves could be the following: let us consider
propagating stationary wave with some given slope of the front, much longer wave can be treated as a presence of
a strong background flow. If the direction of the flow is opposite to direction of wave propagation the slope of
the wave's front will increase. This is what we see in our simulations. Average steepness
$\mu = \sqrt{<|\vec \nabla\eta|^2>}$ has changed: with condensate $\mu_{c}\simeq 0.14$, without condensate
$\mu_{nc}\simeq 0.12$. More detailed picture is given by probability distribution functions (PDFs) for
surface slopes (Fig.~\ref{PDF_grad_eta_x}-\ref{PDF_grad_eta_y_both}). Although maximums of distributions are well described by Gaussian
distribution (which is one of the assumption of the weak turbulence theory), we have significant non-Gaussian tails
and, what is more important, widths of PDFs are different. It means, that in the presence of condensate steeper
waves are more probable.
Surface elevation PDFs, given in Fig.~\ref{PDF_eta} in both cases are in a good agreement with Tayfun
distribution~\cite{Tayfun1980}, which is the first nonlinear correction to Gaussian distribution.
In nature it will result in stronger ``whitecapping'': formation of white foam cap
on the crest of the wave causing additional transport of energy to the small dissipative scale. In the framework of
our model such micro-wavebreaking is impossible. Dissipation in the system prevents formation of strong
spectrum tails corresponding to formation of discontinuities on the surface. Nevertheless, the mechanism is
quite similar: higher steepness means stronger nonlinearity in our system. In this case for harmonics close enough
to the dissipation region generation of second and third harmonics acts as fast and effective additional process of
energy transport to the dissipative scales. Processes, corresponding to multiple harmonics generation are
non-resonant and they are neglected in the theory of wave turbulence. Also it explains why in the experiment in
the framework of Zakharov's equations~\cite{AS2006} spectra were close to weak turbulent. Zakharov's equations
take into account only resonant interactions and do not describe multiple harmonics generation.
We can see, that catastrophic events, like
formation of sharp crests, which cannot be described in the statistical framework of kinetic equation, can
significantly affect physics in the system. The waves kinetic equation can be augmented by additional
dissipation term to simulate this dissipation. As it was shown in recent open field~\cite{BBY2000}
and numerical~\cite{ZKPR2007} experiments, whitecapping dissipation is a phenomenon similar to a second order phase
transition, so even such a moderate change of the average steepness as we observed can cause significant altering
of the energy transfer mechanism. Our results in the presence of condensate can be considered as the first
proof of a conjecture~\cite{NZ2008}, that Phillips spectrum corresponds to a physical picture when a
balance between nonlinear transport terms and intermittent dissipation takes place.

{\it --Conclusion --}
In this Letter author presented results of the first direct numerical simulation
of the direct cascade in the presence of inverse cascade and condensate. The importance of condensate
as a factor, which increases average steepness and stimulates additional intermittant dissipation, is demonstrated.
Qualitative explanation of observed spectra is given. The quantitative explanation is a subject of
further investigations. Still there is no comprehensive theory of whitecapping, which includes analysis of fully
nonlinear equations. One can use presented results for explanation of observed differences in the spectra
in open see and water tanks experiments.

\begin{acknowledgments}
Author would like to thank V.\,E.~Zakharov, V.\,V.~Lebedev, and I.\,V.~Kolokolov for fruitful discussions.

This work was partially supported by RFBR grant 06-01-00665-a,
the Program ``Fundamental problems of nonlinear dynamics'' from the RAS
Presidium and ``Leading Scientific Schools of Russia'' grant
NSh-7550.2006.2.

The author would also like to thank creators of the
open-source fast Fourier transform library FFTW~\cite{FFTW}.
\end{acknowledgments}

\end{document}